\documentclass[aps,prb,superscriptaddress, twocolumn, nofootinbib]{revtex4-1}
\usepackage{braket}						   % braket				
\usepackage{blindtext}
\usepackage{bm}
\usepackage{amsmath,amssymb,amsfonts,latexsym,fancyhdr,graphicx,times}
\usepackage{simplewick}
\usepackage[english]{babel}
\usepackage{graphicx}
\usepackage{subfigure}
\usepackage{xcolor}
\usepackage{color}
\usepackage{verbatim}
\usepackage[pdftex]{hyperref}
\usepackage{enumitem}
\usepackage{mathtools}

\renewenvironment{widetext@grid}{%
	\par\ignorespaces
	\setbox\widetext@top\vbox{%
		\vskip15\p@
		\hb@xt@\hsize{%
			\leaders\hrule\hfil
			\vrule\@height6\p@
		}%
		\vskip6\p@
	}%
	\setbox\widetext@bot\hb@xt@\hsize{%
		\vrule\@depth6\p@
		\leaders\hrule\hfil
	}%
	\onecolumngrid
	\let\set@footnotewidth\set@footnotewidth@ii
}{%
	\par
	\twocolumngrid\global\@ignoretrue
	\@endpetrue
}%

\begin{document}
\title{Study of the energy variation in many-body open quantum systems: 
		\\role of interactions in the weak and strong coupling regimes.}

\author{N.W. Talarico}
\affiliation{QTF Centre of Excellence, Turku Centre for Quantum Physics, Department of Physics and 
Astronomy, University of Turku, 20014 Turku, Finland}
\author{S. Maniscalco}
\affiliation{QTF Centre of Excellence, Turku Centre for Quantum Physics, Department of Physics and 
Astronomy, University of Turku, 20014 Turku, Finland}
\affiliation{ QTF Centre of Excellence, Department of Applied Physics, Aalto University, FI-00076 Aalto, Finland}
\author{N. Lo Gullo}
\affiliation{QTF Centre of Excellence, Turku Centre for Quantum Physics, Department of Physics and 
Astronomy, University of Turku, 20014 Turku, Finland}

\begin{abstract}
We derive an expression for the rate of change of the energy of an interacting many-body system connected to macroscopic leads. We show that the energy variation is the sum of contributions from each different lead. Unlike the charge current each of this contribution can differ from the rate of change of the energy of the lead. We demonstrate that the discrepancy between the two is due to the direct exchange of energy among the considered lead and all other leads. We conclude that the microscopic mechanism behind it are virtual processes via the interacting central region. We also speculate on what are the implications of our findings in the calculation of the thermal conductance of an interacting system. \\
\end{abstract}
\selectlanguage{english}

\maketitle
\section{Introduction}
Understanding how quantum systems conduct energy is currently at the centre of investigation in different branches of physics.
Recent advances in the manipulation of nanoscale systems~\cite{jukka2009} have given a new boost 
to theoretical investigations in that direction; nevertheless how energy and most intriguingly heat 
propagates through a medium has always been at the centre of theoretical debate.
A better understanding of the energy exchange mechanism is also crucial to design devices which can act as (local) refrigerators~\cite{mikko2018}, energy harvesters~\cite{jaliel2019} or can be used as a heat valve~\cite{jukka2018}. The study of energy currents at nanoscale also helps in the characterization of certain physical effects such as Seebeck effect~\cite{svilans2018,clemens2019}, violation of the Wiedemann-Franz law~\cite{bivas2017} and rectification
of currents which are useful in applications at nanoscales~\cite{giazotto2015}. Another interesting problem is understanding the connection between the diffusion properties in a closed system and 
its conduction properties when it is connected to external leads (open system)~\cite{dhar2018, dhar2019}. 
This question also arises in the case of transport in classical systems~\cite{li2003}
because it is known that the transport properties are modified by the nature of the scattering and the type of reservoirs chosen.

The Landauer-B\"uttiker formula for the particle current through a noninteracting system is a milestone in transport theory~\cite{landauer1957,buttiker1986}. It was readily extended to the case of interacting systems by Meir and Wingreen~\cite{Meir1992,Jauho1994} giving the possibility of studying the effects of interactions on the transport properties of many-body quantum systems. It relates the current to the spectral features of the system, therefore giving a microscopic picture of the mesoscopic transport properties. Specifically, in the original work the transport features of the single-impurity Anderson model (SIAM) are linked to the spectral function of the system coupled to the leads in both the Coulomb blockade and Kondo regimes.
The Meir-Wingreen expression, which holds in the stationary state, has also been extended to transient times allowing us to study transport in driven systems~\cite{riku2013, Haugbook, stefleebook, riku2016,riku2018}.
The energy current is defined as the variation of the Hamiltonian of the lead; in this case one obtains an expression analogous to the Meir-Wingreen one for particle current, which holds for both interacting and non-interacting systems provided that the leads are assumed to be non-interacting. In the case of transport through an interacting central region, these expressions correctly capture the many-body interactions through the transmission function or alternatively through the spectral density of the central correlated region~\cite{wang2014}.
A strictly related issue is the definition of heat current, namely the contribution to energy current which is irreversibly dissipated into the lead and that is consistent with the laws of thermodynamics.
The bone of contention is the coupling energy term which needs to be assigned either to the system, or to the terminal, or  be split among the two~\cite{esposito2015,misha2014,misha2016,misha2018, eich2014, eich2016, lopez2015}. The debate about the definition of heat current, although interesting and very important, goes beyond the purpose of this work. \\
In our work, we are interested in the energy variation of the central correlated region due to the reservoirs and we study the difference of this quantities with the energy variation of the reservoirs itself. We use the the non-equilibrium Green's function formalism which naturally allows us to account for both many-body interactions and coupling to external reservoirs through the self-energy functional. We show that the two energy variation need not to be equal (and opposite in sign) as it is for particle currents; we discuss this discrepancy by looking at a specific example: the single-impurity Anderson model.

\section{Non-equilibrium Green's function approach} 
\begin{figure}[t]
	\begin{center}
		\includegraphics[width=.5\textwidth]{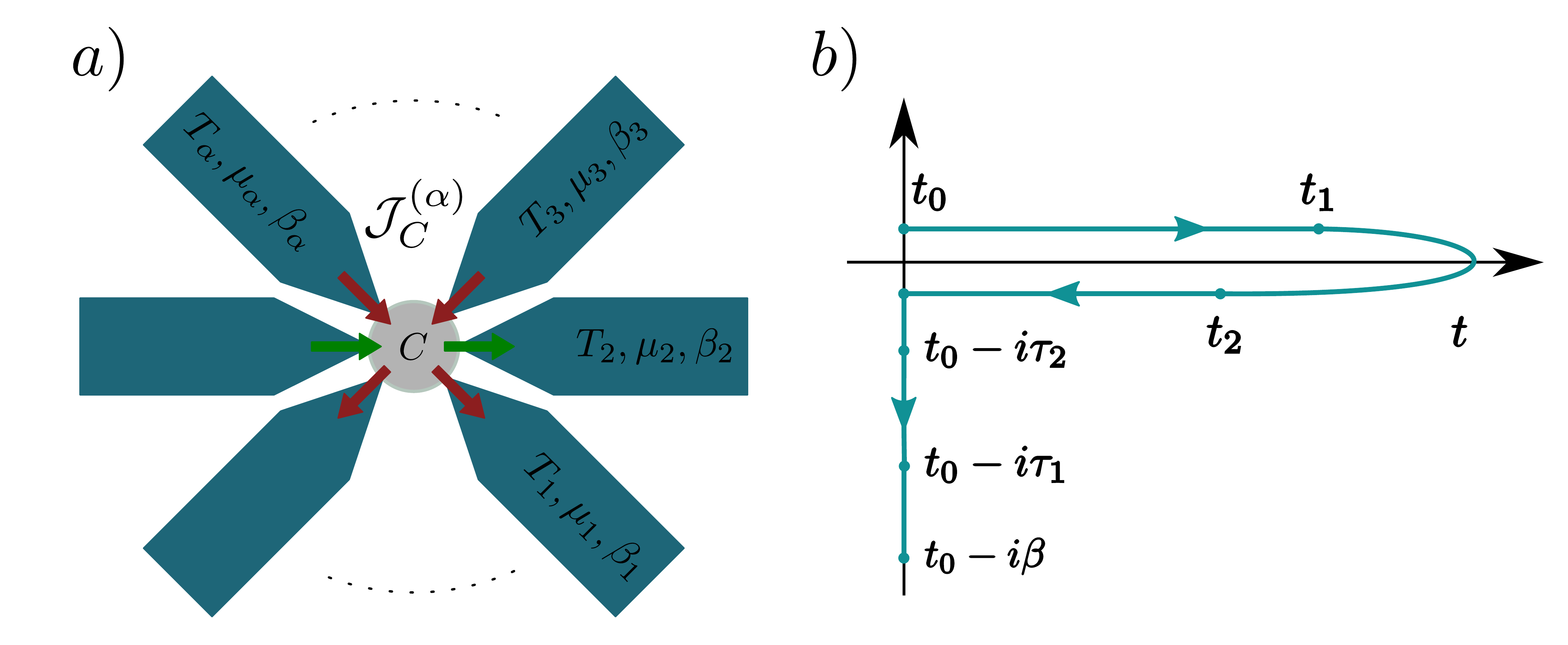}
		\caption{ (Color online) a) Schematic representation of a central system coupled with $N$ leads each characterized by its own chemical potential $\mu_\alpha$ and inverse temperature $\beta_\alpha$. Particle (green arrows) and energy (red arrows) currents flow across the central region due to the biased leads. b) The Keldysh-Schwinger contour $\gamma$. The arrows show the time ordering of the arguments along the complex contour.  }
		\label{fig:sys}
	\end{center}
\end{figure}

We consider a correlated central region where particles can interact through a two-body potential connected to $N$ leads where particles are assumed to be non-interacting. This is the only assumption we make and it is the same needed to derive the Meir-Wingreen expression (and its time dependent version). A schematic representation of the system we consider is shown in  Fig.~\ref{fig:sys} panel $ a)$. 
The Hamiltonian of the total system is given by $\hat{H}(z) = \hat{H}_C(z) + \sum_{\alpha=1}^{N} \hat{H}_{\alpha}(z) + \sum_{\alpha=1}^{N}\hat{V}_{C}^{(\alpha)}(z)$, where $z$ is a generic complex time variable on the Keldysh contour $\gamma$~\cite{Haugbook, stefleebook,bonitzbook,Keldysh1965} (Fig.~\ref{fig:sys} panel $ b )$).
In the central region the Hamiltonian reads $\hat{H}_C(z) = \int d \bold{x}_1\; \hat{\psi}^\dagger (1) h(1) \hat{\psi} (1)+\frac{1}{2}\int d \bold{x}_1 d1'  \;\hat{\psi}^\dagger (1) \hat{\psi}^\dagger (1') v(1, 1') \hat{\psi} (1') \hat{\psi} (1)\nonumber$. \\
The indices $1= (\bold{x}_1, z_1)$, $1'= (\bold{x}_1', z_1')$ are collective indices for the position-spin coordinate  $\bold{x}= (\bold{r},\sigma) $ and complex time $z$, $h(1)$ the single particle Hamiltonian in the central region and  $v(1,2)= \delta_\gamma(z_1-z_2) v(\bold{x}_1,z_1; \bold{x}_2, z_2)$ a generic two-body interaction.
$\hat{H}_\alpha (z)= \int d \bold{x}_1 \hat{\psi}^\dagger_\alpha (1)  h_\alpha(1)\hat{\psi}_\alpha (1)$ describes particles in the region of the $\alpha$-th lead with $h_\alpha(1)$ the single particle Hamiltonian.
The Hamiltonian accounting for the coupling between the interacting region and the leads is chosen to be tunnel-like and given by $\hat{V}_{C}^{(\alpha)}(z) = \int d \bold{x}_1 \left( \hat{\psi}^\dagger (1) T_{\alpha}(1) \hat{\psi}_\alpha (1)+ \text{h.c.}\right)$,
where $T(1)$ are the tunneling energies between particles in the interacting region and those in the $\alpha$-th lead.
The key object in the NEGF framework is the single-particle Green's function (SPGF) $ G(1,1') = -i \left\langle \mathcal{T}_\gamma \hat{\psi}(1) \hat{\psi}^\dagger(1') \right\rangle_0$
with $\mathcal{T}_\gamma$ the time-ordering operator over the Keldysh contour $\gamma$ and $\left\langle \dots \right\rangle_0$ the average over the initial many-body state. The knowledge of $G$ gives access to all single particle quantities (e.g. density, momentum distribution, density-of-states).
For an interacting system the SPGF can be found by solving either the Kadanoff-Baym equations~\cite{stefleebook,bonitzbook,robert2009,robert2016}, the Dyson equation~\cite{logullo2016,talarico2019}, or other equivalent techniques~\cite{spicka1986,bonitz2012,eckstein2014,daniel2018}. A crucial point is the choice of a self-energy, which is typically a functional of the SPGF itself. The reason to introduce the self-energy is to truncate the hierarchy of equations which couples the equation-of-motion for the SPGF to those of higher-order Green's functions~\cite{stefleebook}.
Henceforth, we shall assume that a self-energy, accounting for both the many-body interactions and the coupling to the leads, has been chosen and that a solution for the SPGF has been found. Under these assumptions we shall compute the energy variation in the interacting region in a way which is consistent with the approximations embodied in the chosen self-energy.

\section{Energy variation in the central region}
We start considering the rate of variation of the energy in the central (interacting) region $d\langle \hat{H}_{C}(t) \rangle/ dt = i \langle [\hat{H} (z),\hat{H}_C (z)] \rangle_{z=t} +   \partial\langle\hat{H}_C(z) \rangle/\partial z|_{z=t}\equiv\mathcal{J}(t)+P(t)$ where we defined the energy current $\mathcal J(t)$ and the power $P(t)$ due to external drive; in the following we will disregard this last term as it is not relevant for the purposes of our discussion. By computing the commutator of $\hat H_C(z)$ with the total Hamiltonian~(see Appendix \ref{app:energy} for details), we obtain an expression in terms of $G_{C \alpha}(1,1') = -i \langle \mathcal{T}_\gamma \hat{\psi} (1) \hat{\psi}_\alpha^\dagger ( 1') \rangle$  and $G^{(2)}_{CCC \alpha}  (1,2;3,4) = (-i)^2 \langle  \mathcal{T}_\gamma \hat{\psi}( 1) \hat{\psi} ( 2) \hat{\psi}_\alpha^\dagger (4) \hat{\psi}^\dagger (3) \rangle$.\\
This expression can be manipulated by means of the S-matrix expansion in the interaction picture with respect to $\hat H_C(t)+\sum\limits_\alpha\hat H_\alpha(t)$ to obtain $\mathcal{J}(t)=\sum\limits_\alpha\mathcal{J}_C^{(\alpha)}(t)$~(see Appendix \ref{app:mixed} for details) with:
\begin{align}
\label{eq:ecurrent2}
&\mathcal{J}_C^{(\alpha)}(t)=2 \text{Re} \left\{ \int d \bold{x}_1 d \bar{1} d \bar{2}\; H(1,\bar{1})  G (  \bar{1}, \bar{2} ) \Sigma_{ \alpha} (\bar{2},\bar{1}^+)    \right\}_{z_1=t}
\end{align}
where $H(1,1')= \left[ h(1)\delta(1,1')+  \Sigma (1,1') \right]$  and $1^+=(\bold{x}_1, z_1^+)$, with $z_1^+$ a time infinitesimally later than $z_1$ on the Keldysh-contour.\\
Eq.~ \eqref{eq:ecurrent2} expresses the variation of the energy in the interacting region as sum of different contributions each coming from the individual leads.
The single contribution from the lead $\alpha$ is in turn made of two terms; the first one is similar to the expression of the particles' current due to the $\alpha$-th lead: $\mathcal{I}_C^{(\alpha)}(t)= 2 \text{Re} \left\{ \int d \bold{x}_1 d \bar{1} d \bar{2}\;  G (  \bar{1}, \bar{2} ) \Sigma_{ \alpha} (\bar{2},\bar{1}^+)\right\}_{z_1=t}
$, with the significant difference that it contains the single-particle Hamiltonian $h(1)$ and it accounts for the energy carried by the flowing particles; the effect of the interactions is present only in the SPGF $G$ through a change in the density of states of the particles in the central region. This term is analogous to the energy variation of the lead where the single particle Hamiltonian of the central region is replaced by the single particle Hamiltonian of the $\alpha$-th lead.
The second term contains the many-body self-energy explicitly and therefore accounts for the transport of the particle-particle interaction energy; it can be seen as a mechanism of energy redistribution in the central region due to particle-particle scattering: non-interacting electrons coming from the leads scatter in the central region releasing part of their energy before tunneling into a new non-interacting state of another lead.

\section{Inter-leads coupling} 
\label{sec:leads}
The expression in Eq.~\eqref{eq:ecurrent2} has to be compared with the Meir-Wingreen like expression $J_\alpha(t)=i\langle[\hat H(z),\hat H_\alpha(z)]\rangle_{z=t}$, that describes the energy variation of the lead $\alpha$ due to the coupling with the central correlated region. For the particle currents, it is true that the current into the lead is equal and opposite to the charge variation in the central system due to that lead, namely $\mathcal{I}_\alpha=-\mathcal{I}_C^{(\alpha)}$. Here we defined the current in the $\alpha$ lead as $\mathcal{I}_\alpha=i\langle[\hat H(z),\hat N_\alpha(z)]\rangle_{z=t}$ and the currents into the central region due to the lead $\alpha$ as $\sum\limits_\alpha \mathcal{I}_C^{(\alpha)}=i\langle[\hat H(z),\hat N_C(z)]\rangle_{z=t}$. This relations translates into formal mathematical expression the intuitive concept of the locality of the number operator. Basically, they state that particles flowing out of a lead necessarily enter the central region causing an equally and opposite variation.
In analogy with the particle current we would expect that the following relation for the energy current $\mathcal{J}_\alpha(t)=-\mathcal{J}_C^{(\alpha)}(t)$ should always be true as well. However, as we prove in the following, this is not always the case. It is sufficient to look at the time derivative $d \langle \hat{V}_{C}^{(\alpha)}(t) \rangle/dt$ of the coupling Hamiltonian between the terminal $\alpha$ and the central region. It is easy to prove~(see Appendix \ref{app:coupling} for details) that 
\begin{equation}
 \label{eq:ccurrent}
 \frac{d}{dt} \langle \hat{V}_{C}^{(\alpha)}(t) \rangle = -\mathcal{J}_\alpha(t)-\mathcal{J}_C^{(\alpha)}(t)-\Delta \mathcal{J}_{\alpha}(t)+ \frac{\partial}{\partial t}\langle\hat{V}_{C}^{(\alpha)}(t) \rangle,
\end{equation}
where $\Delta J_{\alpha}(t)= \sum\limits_{\beta\neq \alpha}\int d \bold{x}_1 T_\beta(1)G_{\beta\alpha}(1;1^+)  T_\alpha^*(1)$ and $G_{\beta \alpha}(1,1') = -i \langle \mathcal{T}_\gamma \hat{\psi}_\beta (1) \hat{\psi}_\alpha^\dagger ( 1') \rangle$ is the  lead $\beta$- lead $\alpha$ Green's function. It has been shown~\cite{misha2016} that the DC component of $\frac{d}{dt} \langle \hat{V}_{C}^{(\alpha)}(t) \rangle=0$ and therefore $\mathcal{J}_\alpha(t)=-\mathcal{J}_C^{(\alpha)}(t)-\Delta \mathcal{J}_{\alpha}(t)$. 
Therefore, we have shown that a variation of the energy of the lead is not necessarily accompanied by an equal and opposite change in the energy of the central region as in the case for the particle variation. The difference between the two is given by the term $\Delta \mathcal{J}_\alpha(t)$ which arises from the direct propagation of particles from any other lead $\beta(\ne\alpha)$ and the lead $\alpha$ via virtual scattering through the central region. This interpretation is supported by the physical meaning of $G_{\beta \alpha}(1,1')=
\int d \bar{1}d \bar{2}\;g_{\beta} (1 ;\bar{1}) T_\beta^*(\bar{1}) G (\bar{1};\bar{2})  T_\alpha(\bar{2})  g_{\alpha} (\bar{2} ;1')$~(see Appendix \ref{app:coupling} for details): it represents the propagation of a particle from one lead to another through the central region. The expression $\Sigma_{\beta\alpha}(1;1')=T_\beta^*(1) G (1;1')  T_\alpha(1')$ is also called the inbedding self-energy~\cite{stefleebook} and accounts for the back-action of the central region onto the leads. 
This term is usually either negligible or zero altogether. In the weak coupling limit this term is vanishingly small due to the fact that it is of fourth order in the coupling between the central region and the leads. Moreover it also vanishes when the two integrals in  the expression for $G_{\beta\alpha}(1;1')$ have different spatial supports, namely when $T_\alpha(1)$ and $T_{\beta}(1')$ for $\alpha\neq\beta$ are non-zero on different spatial regions and therefore their product vanishes.

Nonetheless, it is possible to envisage situations in which this term gives a non-vanishing contribution: short quantum wires (mean free path comparable with the size of the wire), more than one lead coupled to the same spatial region or spatially extended coupling between the leads and the central region.
As a consequence, the energy variation of the lead is not only related to the one of the central region, as one would expect, but also accounts for contribution coming from the direct exchange of energy with other leads. Therefore, as one can easily deduce, it is crucial to consider this term in situations where one wants to infer the thermal transport properties of a system (the central region) by measuring the properties of the reservoirs (e.g. particle distribution, temperatures).

\section{SIAM in the Kondo regime} 
In order to show the consequences of the results presented so far we look at the energy-transport in the single-impurity Anderson model~\cite{anderson,glazman2001} described by the Hamiltonian $\hat{H} = \varepsilon \sum_\sigma \hat{d}_\sigma^\dagger \hat{d}_\sigma + U \hat{d}_\uparrow^\dagger \hat{d}_\uparrow \hat{d}_\downarrow^\dagger \hat{d}_\downarrow +\sum_{\alpha, k \sigma} (\varepsilon_ {\alpha k \sigma} + \mu_\alpha) \hat{c}_{\alpha k \sigma}^\dagger \hat{c}_{\alpha k \sigma} \nonumber-g \sum_{\alpha, k \sigma} ( \hat{c}_{\alpha k \sigma} \hat{d}_{\sigma}^\dagger + h.c. )$.
Here, $\hat{d}_{\sigma}^\dagger\left(\hat{d}_{\sigma}\right)$ corresponds to creation (annihilation) of electron on the impurity level with spin $\sigma$, $\varepsilon$ denotes the single-particle energies, and $U$ is the electronic charging energy. The operator $\hat{c}_{\alpha k \sigma}^\dagger\left(\hat{c}_{\alpha k \sigma}\right)$ creates (annihilates) electron with state $k$ and spin $\sigma$ in the lead $\alpha=L,R$ with chemical potential $\mu_\alpha$. Finally, $g$ denotes the tunneling amplitude between the terminal and the impurity level. In the wide-band limit approximation we are left with a frequency-independent coupling $\Gamma = |g|^2$.
In what follow we consider unbiased voltage ($\mu=\mu_L=\mu_R=0 $), a finite and symmetric thermal bias $T_L = T+ \Delta T/2$, $T_R = T- \Delta T/2$, with $T=(T_L +T_R)/2 = 0.125$ and $\Delta T=T_L -T_R = 0.15$ and we study the dynamical quantities as function of the impurity level position $\varepsilon$. We present our results as a function of the shifted single-particle energies or gate-voltage $v_g = \varepsilon + U/2$, such that the particle-hole symmetric point $\varepsilon=-U/2$ correspond to $v_g=0$. \\
Under certain conditions the SIAM may reveal Kondo effects characterized by the formation of a correlated asymmetric resonance in the spectral structure of the system \cite{haldane, glazman2001}. The conditions to attain such regime are a large charging energy and tunnel coupling at low temperature, $U \gg \Gamma \gg T$ and $T \ll T_K $. The expression for the threshold temperature or Kondo temperature is given by  
$T_K = \frac{1}{2} \sqrt{ \Gamma U} exp \left( \pi \frac{(v_g^2 - U^2/4)}{\Gamma U} \right) $  \cite{hewson}  and is strictly valid in the Kondo regime where $-U/2+ \Gamma/2 <v_g< U/2- \Gamma/2$. 
\begin{figure}[t]
	\begin{center}
		\includegraphics[width=.5\textwidth]{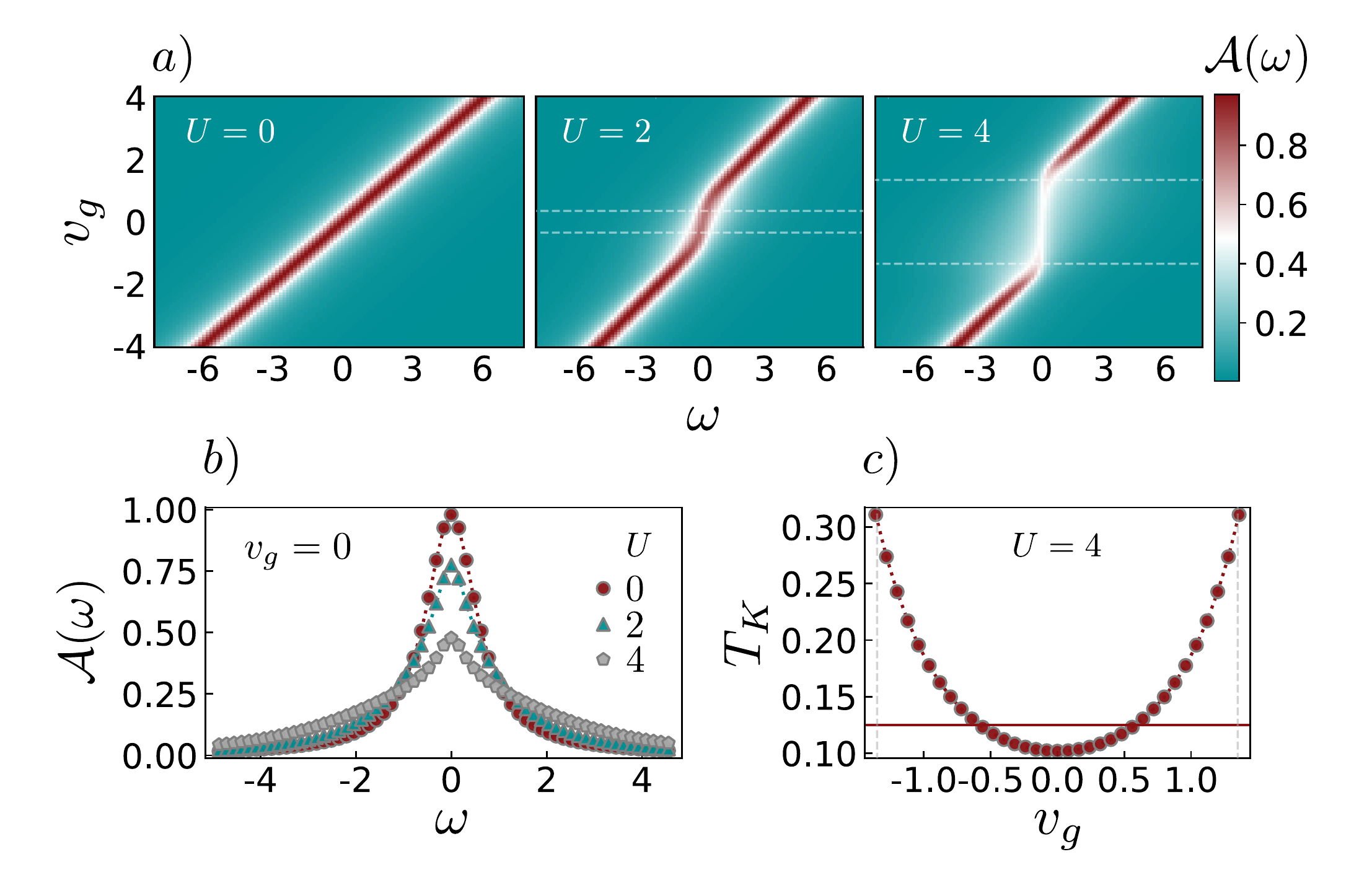}
		\caption{Characterization of the Kondo regime. $a)$ Density plot in the strong coupling regime $\Gamma=1.3$ and different interactions strength $U=0,2,4$ of the non-equilibrium spectral function $ \mathcal{A}(\omega)$ in the gate-voltage/frequency plane $v_g- \omega$. The white dashed lines represent the range of gate-voltage where the Kondo correlations are expected $-U/2+ \Gamma/2 <v_g< U/2- \Gamma/2$ for the respective interactions strength. $b)$ Corresponding non-equilibrium spectral function at the particle-hole symmetric point $v_g=0$ , $\varepsilon=- U/2$ for the different interactions considered. It is possible to appreciate how the shape of spectral function changes as the Kondo regime is reached $U/\Gamma \gg 1$. $c)$ Kondo temperature $T_K$ as a function of $v_g$ in the range of gate-voltage where Kondo correlations occur for the charging energy $U=4$. The red solid line represents the value of the "average" temperature of the leads $T=(T_L +T_R)/2 = 0.125$ . }
		\label{fig:kondo}
	\end{center}
\end{figure}
The interactions in the correlated impurity are treated in the self-consistent $GW$ approximation (see Appendix \ref{app:gw} for details), which on the one hand is guaranteed to satisfy macroscopic conservation laws \cite{stefleebook}, and on the other hand allows us to explore certain features of the Kondo regime \cite{rubio2008}.
We solve the Dyson equation numerically using the method described in Ref.~\cite{logullo2016,talarico2019}.
We checked that the obtained expression for the total current through the interaction region $\mathcal{J}(t)$ in Eq.~\eqref{eq:ecurrent2} is consistent with the derivative of the total energy in the corresponding region $d \mathcal{E}(t)/dt$ where $\mathcal{E}(t) =\langle \hat{H}_C (t) \rangle = \left[ i \int d \bold{x}_1   d 2 \;\biggl[  h(1) \delta(1,2)  +\frac{1}{2}\Sigma(1,2) \biggr] G (2,1^+)\right]$.
We find, within numerical errors, a remarkable agreement between these two quantities at any time of the dynamics and for any set of the parameters considered (Appendix \ref{app:gw}).
This demonstrates that the approximations done on the two-particle Green's function, while working out an expression for $\mathcal{J}(t)$, are indeed consistent with those used to solve the Dyson equation for the single-particle Green's function and embodied into the chosen self-energy.

\begin{figure*}[t]
	\begin{center}
		\includegraphics[width=1.0\textwidth]{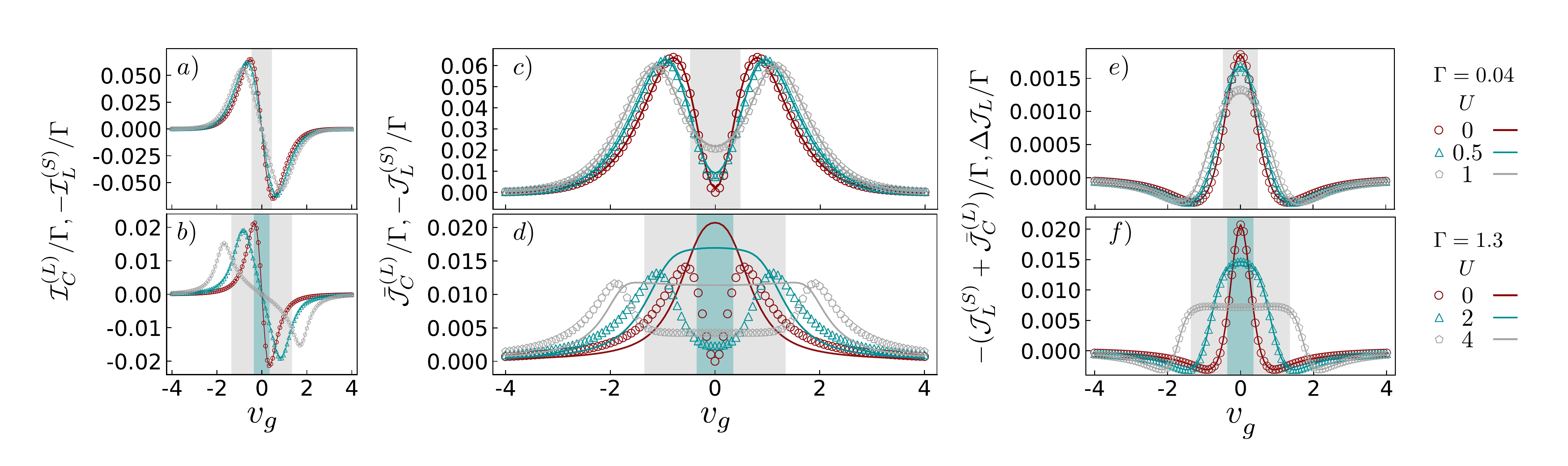}
		\caption{Particle and Energy currents. $(a)-(b)$ comparison of the steady state particle currents flowing through the central region due to the left lead $\underset{t\rightarrow\infty}{\lim}\; \mathcal{I}_C^{(L)}(t)$ (markers) and the reverse current that runs into the same lead $-\mathcal{I}_L^{(S)}$ (solid lines) in the weak coupling  $\Gamma= 0.04$ $(a)$ and in the strong coupling regimes $\Gamma= 1.3$ $(b)$ at the charging energy  $U$ as indicated in the legend. $(c)-(d)$ parallel between the steady state energy currents across the interacting system from the left lead $\bar{\mathcal{J}}_C^{(L)}$ (markers)  and the opposite of the energy current flowing into the respective lead  $-\bar{\mathcal{J}}_L^{(S)}$ (solid lines) at the same couplings and charging energies as in the case of particle currents. $(e)-(f)$ comparison between (minus) the sum of the two steady state energy currents $-(\bar{\mathcal{J}}_L^{(S)}+\bar{\mathcal{J}}_C^{(L)}  )$ (markers)  and the term $\underset{t\rightarrow\infty}{\lim}\; \Delta\mathcal{J}_L(t)$ (solid lines) in the region of parameters as specified before. All the quantities are normalized by their respective $\Gamma$. The shade areas in the figures correspond to the gate-potential range of the Kondo regime $-U/2+ \Gamma/2 <v_g< U/2- \Gamma/2$ for the related couplings and charging energies.  }
		\label{fig:compw}
	\end{center}
\end{figure*}

Before moving to the systematic study of the currents, we first characterize the regimes of parameters where the Kondo effects may occur. Fig.~\ref{fig:kondo} panel $ a)$ shows a map in the gate-voltage/frequency plane of the non-equilibrium spectral function of the system $ \mathcal{A}(\omega) = i [G^R(\omega) -G^A (\omega)]$  obtained from the solution of the Dyson equation in a strong coupling regime with the leads  $\Gamma=1.3$ and for different charging energies $U=0,2,4$. As the interaction increases, one can identify a range in the gate-voltage where the spectral function is centered at the chemical-potential of the leads, that correspond exactly to the range where Kondo correlations are expected $-U/2+ \Gamma/2 <v_g< U/2- \Gamma/2$ (white dashed lines). Fig.~\ref{fig:kondo} panel $ b)$ shows instead how the Kondo correlations reshape the spectral function at the charge degeneracy point $v_g=0$. Going from zero to large interactions, the Lorentzian shape is lost to leave a more asymmetric and pinned structure.\\ 
Fig.~\ref{fig:kondo} panel $ c)$ shows the Kondo temperature $T_K$ for the charging energy $U=4$ as a function of $v_g$ in the range of gate-voltage where Kondo correlations take place. As we can see the average temperature $T$ (red solid line) lies slightly above the Kondo temperature $T_K$ in a region close to the degeneracy point $v_g=0$. Nonetheless, the spectral function is still pinned at the chemical potential of the leads in the whole interval and its shape is very different from a Lorentzian, a clear signature of the Kondo regime. This may be explained from the highly out-of-equilibrium scenario ($T \sim \Delta T$) that we consider; it is quite reasonable that the system does not equilibrate to the average temperature $T$ of the leads and instead reaches a lower temperature that allows the Kondo correlations to persist.

\section{Particle and energy currents}
We now look at the steady-state properties of the time-dependent expressions for the particle and energy currents, namely the limit $t \to \infty$.
In the case of two symmetrically coupled terminals the particle current is given by the Meir-Wingreen formula $\underset{t\rightarrow \infty}{\lim}\mathcal{I}_\alpha(t)\equiv\mathcal{I}_\alpha^{(S)} = \int d \omega \Gamma (\omega) [f_\alpha (\omega) - f_{\bar{\alpha}} (\omega) ]  \mathcal{A}(\omega)$. Here $\Gamma (\omega)= \Gamma_L (\omega) \Gamma_R (\omega)/[\Gamma_L (\omega)+\Gamma_R (\omega)] $, $\bar{\alpha} \ne \alpha $, $f_\alpha (\omega)= (1+ e^{\beta_\alpha(\omega - \mu_\alpha)})^{-1}$ is the Fermi-Dirac distribution function for electrons in the lead $\alpha$ and $ \mathcal{A}(\omega)$ is the non-equilibrium spectral function of the central region.
For the energy current we compare $\bar{\mathcal{J}}_C^{(L)}\equiv\underset{t\rightarrow \infty}{\lim}\mathcal{J}_C^{(L)}(t)$ with $\underset{t\rightarrow \infty}{\lim}\mathcal{J}_L(t)\equiv\mathcal{J}_L^{(S)} = \int d \omega \;\omega \Gamma (\omega) [f_\alpha (\omega) - f_{\bar{\alpha}} (\omega) ]  \mathcal{A}(\omega)$.
The latest expression is the energy counterpart of the Meir-Wingreen formula for the particle current and properly describes how the energy runs across the non-interacting leads. It is widely used to characterize thermoelectric properties of correlated materials in linear response ~\cite{kim2002,lei2002,costi2010}.
Fig.~\ref{fig:compw} shows the agreement between the particle current in the central region due to the left lead $\mathcal{I}_C^{(L)}(t), (t \Gamma\gg 1)$ and the opposite of the current that flows into the same lead $-\mathcal{I}^{(S)}_L$ for both weak coupling $\Gamma \ll \Delta T$ panel $a)$ and  strong coupling $\Gamma \gg \Delta T$ panel $b)$, for different values of the interaction strength $U$. The two currents are actually equal and opposite regardless the regime consider and for any range of parameters chosen as one would expected.
We observe that for positive gate voltages particles flow from the central region to the left lead $\left(\mathcal{I}_C^{(L)}(t)=-\mathcal{I}^{(S)}_L<0\right)$ because the position of the spectral function of the system lies above the chemical potential of the lead ($\mu_L=0$). The opposite situation is realized for negative gate potentials where particles flow from the left lead to the central region $\left(\mathcal{I}_C^{(L)}(t)=-\mathcal{I}^{(S)}_L>0\right)$.
In the weak-coupling case,  panel Fig.~\ref{fig:compw} $a)$, the particle current is fully specified by the window function of the leads  $\mathcal{I}_L \sim \Delta f (v_g)=f_L(v_g)-f_R(v_g)$. This may be understood from the shape of the spectral function that, although being a Lorentzian function, is much sharper and narrow than the energy window of the leads ($\Gamma \ll \Delta T$) and it can be approximated with $\mathcal{A}(\omega) \sim \delta(\omega - v_g)$. 
This picture is slightly different when the interaction is present since, in this case, the overall effect is to broaden the spectral function that cannot anymore be treated as a delta function. 
Nonetheless, the physical scenario is qualitatively the same and only quantitatively different. 
The strong-coupling case, Fig.~\ref{fig:compw} $b)$, does not present significant differences outside of the Kondo region. When the system enters the Kondo regime (green and gray shaded areas) the behavior of the current is drastically different from the window function $\Delta f(\omega)$ reflecting the change in the shape of the spectral function in this regime. \\ 
In contrast to the particle current, a perfect agreement between the energy current across the interacting region due to the left reservoir $\bar{\mathcal{J}}_C^{(L)}$ and the opposite current in the same lead $-\mathcal{J}_{L}^{(S)}$ is found only in the weak-coupling scenario as shown in Fig.~\ref{fig:compw} panel $c)$. At strong coupling we observe a qualitatively different behavior and not only a merely quantitative deviation between this two expressions, Fig.~\ref{fig:compw} panel $d)$, as it was for the particle currents.
Interactions are not responsible for it as the effect is present, and actually more pronounced, in the non-interacting case $(U=0)$. 
The maximum difference between the two currents is at particle-hole symmetric point $(v_g=0)$ where $\bar{\mathcal{J}}_C^{(L)}$ is zero and $\mathcal{J}_{L}^{(S)}$ has its maximum. This can be explained with the following argument: the central region is at resonance, it is completely transparent and all the energy flows from one lead to the other without affecting the energy of the central region. 
The presence of Kondo correlations manifest themselves as a plateau in the energy currents in the range of gate-voltage where they are predicted, namely for $-U/2+ \Gamma/2 <v_g< U/2- \Gamma/2$. These regions are shown in Fig.\ref{fig:compw} as green and gray shaded areas and they corresponds to the regions in which the spectral function is pinned at the chemical potential of the leads, as shown by the horizontal solid lines in Fig.\ref{fig:kondo} panel a). One interesting observation is that the extension of the plateaux of the lead energy current overestimates this regions, whereas the one in the energy variation of the central region shows a better agreement.
As we noticed above the pinning of the spectral function is an hallmark of the  emergence of the Kondo bound state and it can be probed through transport experiments~\cite{martin2018}. In these experiments properties of the central region are inferred from measurements onto the leads. We have just shown that in the determination of the extent of the Kondo regime in gate voltage, it might be beneficial to consider this extra term to get a more precise estimation of the Kondo region for single level quantum dots in the strong coupling regime.
Interestingly, in the case of a thermal bias in the leads the plateau induced by the Kondo cloud appears in the energy current, whereas  in the case of small bias voltage it appears in the particle current as shown in Ref.~\cite{talarico2019}.
In general interactions increase the change of energy in the central region with respect to the non-interacting case.
Nevertheless, there is a reduction between the energy current in the left lead $\mathcal{J}^{(S)}_L$ and the energy change in the central interacting region predicted by Eq.~\ref{eq:ecurrent2}.

\section{Virtual processes}
In Sec.\ref{sec:leads} we have shown that the energy variation of one lead and the lead's contribution to the variation of the the energy of the central region differ by $\Delta J_{\alpha}(t)$. This term has been shown to arise from the direct coupling of the lead to all other leads via second order processes through the central region.
Moreover, we observe that in the strong coupling regime of the SIAM the two currents are indeed different and that, this effect does not come from the strong correlations present in this regime, but actually it is more pronounced in the non-interacting case. If we compute the term $-\mathcal{J}_L^{(S)}-\mathcal{J}_C^{(L)}$ and compare it with $\Delta\mathcal{J}_L$ computed separately we obtain a perfect agreement as shown in Fig.~\ref{fig:compw} $e)-f)$ for both weak and strong coupling respectively.
Recently, this term has been shown~\cite{josefsson2019} to be useful in computing the energy variation in the lead, that has been also measured in a setup analogous to the one we have considered.
It is worth to notice that, whenever the interaction is present, the energy rate contribution $\mathcal{J}_C^{(L)}(t)$ is never zero when the density of states of the dot is in the energy window of the leads. Thus, the presence of the interaction redistributes the energy inside the dot and forbids from a perfect transparency as the one observed in the non-interacting case. \\
The lead-lead coupling has been already discussed in ~\cite{wang2012,wang2014} with the remarkable difference that there a lead-lead interaction was present from the beginning in the Hamiltonian of the system. Moreover, the effect of such coupling on the energy transfer and its microscopic mechanism has not been discussed.
As a concluding remark, we would like to highlight the importance and the implications of the term  $\Delta\mathcal{J}_{\alpha}$ whenever it is not vanishing. When probing the conduction properties, either electrical or thermal, measurements are performed on the leads. In the case of particle currents the rate of change of particles in the lead equals the change of particles in the central region due to the lead itself. Therefore, measure of the charge current {\it into} the lead is equivalent to measure the change in the particle currents in the central region. Hence it is meaningful to infer the electrical conductivity of the central region from the measured current in the leads.
Nevertheless, when it comes to compute the thermal conductivity of the system in the central correlated region $\kappa = \frac{\partial \dot{\mathcal{Q}}_{L}}{\partial \Delta T}\mid_{\mathcal{I}_L^{(S)}=0}$, with the heat current given by $ \dot {\mathcal{Q}}_{L}=(\mathcal{J}_{L}^{(S)}-\mu_L \mathcal{I}_L^{(S)})$, one has to take extra care on what is the actual interpretation of what it is computed. Indeed, according to Eq.\ref{eq:ccurrent}, at stationarity and in the absence of external drive, one has $ \kappa = -\frac{\partial }{\partial \Delta T} \left(\bar{\mathcal{J}}_{C}^{(L)}-\mu_L\mathcal{I}_C^{L} +\Delta \mathcal{J}_{L} \right)\mid_{\mathcal{I}_C^{(L)}=0}$, 
where we replace $\mathcal{I}_L^{(S)}\rightarrow -\mathcal{I}_C^{(L)}=-\underset{t\rightarrow \infty}{\lim}\mathcal{I}_C^{(L)}(t)$.
The last equation shows that to compute the thermal conductance of the central region, the lead-lead term $\Delta\mathcal{J}_L$ needs to be considered. The term, as we discussed, accounts for the energy transfer through the contacts or terminals, thus in the strong coupling or for spatially extended couplings, what it is actually computed is the interface thermal conductance between tow leads, as if the two were directly coupled via a renormalized tunneling. We conclude that in the case of strong coupling and/or when the coupling of the leads are spatially extended, the usual definition of thermal conductance can be misleading as a figure of merit for the thermal properties of the central region. This is in contrast with the electrical conductance which is, instead, always consistent with the rate of change of the particle currents in the middle region. 
As a concluding remark, we would like to comment on the possibility to define, by analogy with the lead, the heat rate $\dot{\mathcal{Q}}_{C}^{(L)}=\bar{\mathcal{J}}_{C}^{(L)}-\mu_L\mathcal{I}_C^{L}$, namely the heat flowing from the central region to the left lead. However, this definition is in contrast with the thermodynamic formulation of heat as we will show in a forthcoming work.

\section{Conclusions}
We have shown that the energy variation of an interacting system coupled to $N$ leads is equal to the sum of different contributions each related with  the energy flowing from/to the individual leads. Each term is in turn made of two further contributions, a first one accounting for the energy transport due to the flow of particles, and a second one accounting for the particle-particle scattering which redistribute the energy in the interacting region. This expression has been compared to the energy variation into the corresponding lead and we have found that the two are not always equal and opposite in sign as it is in the case of particle current. The difference between these two currents is due to the direct exchange of energy among two of the leads and the microscopic mechanism behind it is the coupling of these leads through virtual processes involving the central interacting region.

\begin{acknowledgments}
NLG akcnowledges P. Erdeman, and M. V. Moskalets and P. Burset Atienza for insightful comments on the expression for the energy current.
The authors acknowledge financial support from the Academy of Finland Center of Excellence
program (Project no. 312058) and the Academy of Finland
(Project no. 287750).
NLG acknowledges financial support from the Turku Collegium for Science and Medicine (TCSM).
Numerical simulations were performed exploiting the Finnish CSC facilities under the Project no. 2000962 ("Thermoelectric effects in nanoscale devices").
\end{acknowledgments}

\appendix

\section{ENERGY VARIATION OF THE CENTRAL REGION}
\label{app:energy}
We discuss here how to compute the expression of the energy variation in the central region:
\begin{equation}
\mathcal{\dot E}_C \equiv \frac{d}{dt}\langle  \hat{H}_C (t)\rangle = i  \langle [\hat{H}(t),\hat{H}_C(t)] \rangle
\end{equation}
which can be recast in a form very similar to the time dependent Meir-Wingreen one for the particle current ~\cite{Meir1992,Jauho1994, riku2013}
in terms of the single-particle Green's function, the embedding self-energies and, in this case, of the many-body self-energy.
Here we show how the energy variation can be expressed in terms of single- and two-particle mixed Green's functions.
In the next subsection we demonstrate how to manipulate these expressions in order to recast them in terms of single-particle quantities of the central region. First, we need to compute explicitly the commutators entering into the definition of the energy current:

\begin{widetext}

\begin{align}
\label{eq:current}
&\mathcal{\dot E}_C(z_1)= i  \langle [\hat{H} (z_1),\hat{H}_D (z_1)] \rangle = i  \langle [\hat{H}_T (z_1),\hat{H}_D (z_1)] \rangle \nonumber \\
&= \sum\limits_\alpha  \biggl(i  \left\langle \left[ \int d \bold{x}_1 \left( \hat{\psi}^\dagger (1) T(1) \hat{\psi}_\alpha (1)+ \hat{\psi}_\alpha^\dagger (1) T^*(1) \hat{\psi} ( 1)\right), \int d \bold{x}_2 \hat{\psi}^\dagger (2) h(2) \hat{\psi} (2) \right] \right\rangle + \nonumber \\ 
&+ i  \biggl\langle \biggl[ \int d \bold{x}_1 \left( \hat{\psi}^\dagger (1) T(1) \hat{\psi}_\alpha (1)+ \hat{\psi}_\alpha^\dagger (1) T^*(1) \hat{\psi} (1)\right),\frac{1}{2} \int_\gamma d z_1' \int \int d \bold{x}_2 d \bold{x}_1' \hat{\psi}^\dagger (2) \hat{\psi}^\dagger (1') v(1',2) \hat{\psi} (1') \hat{\psi} (2) \biggr] \biggr\rangle \biggr) \nonumber \\
&=\sum\limits_\alpha \biggl(-i \int d \bold{x}_1 \; T(1) h(1) \left\langle \hat{\psi}^\dagger (1)  \hat{\psi}_\alpha (1) \right\rangle   +i  \int d \bold{x}_1\; T^*(1) h(1)   \left\langle\hat{\psi}_\alpha^\dagger (1) \hat{\psi} (1) \right\rangle + \nonumber \\
& - \frac{i}{2} \int_\gamma d z_1' \int \int d \bold{x}_1  d \bold{x}_1'\; T(1') v(1,1') (-i)^2 \left\langle \hat{\psi}^\dagger (1) \hat{\psi}^\dagger (1')  \hat{\psi} (1) \hat{\psi}_\alpha (1') \right\rangle + \nonumber  \\
& +\frac{i}{2} \int_\gamma d z_1' \int \int d \bold{x}_1  d \bold{x}_1'\; T(1) v(1,1') (-i)^2 \left\langle \hat{\psi}^\dagger (1) \hat{\psi}^\dagger (1')  \hat{\psi} (1') \hat{\psi}_\alpha (1) \right\rangle + \nonumber  \\  
&-\frac{i}{2} \int_\gamma d z_1' \int \int d \bold{x}_1  d \bold{x}_1'\; T^*(1) v(1,1') (-i)^2 \left\langle  \hat{\psi}^\dagger_\alpha (1) \hat{\psi}^\dagger (1') \hat{\psi} (1')  \hat{\psi} (1) \right\rangle + \nonumber  \\
& +\frac{i}{2} \int_\gamma d z_1' \int \int d \bold{x}_1  d \bold{x}_1'\; T^*(1') v(1,1') (-i)^2 \left\langle  \hat{\psi}^\dagger_\alpha (1') \hat{\psi}^\dagger (1) \hat{\psi} (1')  \hat{\psi} (1) \right\rangle \biggr)\nonumber
\end{align}

where $z$ is the complex time variable on the path $\gamma$ in the complex-time plane
and  $1\equiv (\bold{x}_1,z_1)$, $2\equiv (\bold{x}_2,z_2)$ and $1'\equiv (\bold{x}_1',z_1')$ are multi-indexes for position, spin and complex time. It is understood that at the end we take $z_1=t$ to project the complex time variable onto the real axis.

We can rewrite the first two terms in the following way 

\begin{equation*}
\int d \bold{x}_1\; h(1)\left[ T^*(1) i \left\langle  \hat{\psi}_\alpha^\dagger (1) \hat{\psi} (1) \right\rangle - T(1)  i \left\langle   \hat{\psi}^\dagger (1)  \hat{\psi}_\alpha (1) \right\rangle \right] = 2 Re \biggl\{ \int d \bold{x}_1 \;h(1) T^*(1) G_{C \alpha}(1;1^+) \biggr\} 
\end{equation*}
where we use the definition of the mixed Green's functions: $G_{C \alpha}(1,1') = -i \left\langle  \mathcal{T}_\gamma \hat{\psi} (1) \hat{\psi}_\alpha^\dagger (1')\right\rangle$ and $G_{ \alpha C}(1;1')= -i \left\langle  \mathcal{T}_\gamma \hat{\psi}_\alpha (1) \hat{\psi}^\dagger (1')  \right\rangle$, with $\mathcal{T}_\gamma$ the time-ordering operator over the Keldysh contour $\gamma$ in the complex-time plane and where we exploit the property $ G_{ \alpha C}(1;1') = - [G_{C \alpha}(1';1)]^*$. Moreover $1^+\equiv (\bold{x}_1,z_1^+)$ with $z_1^+$ being a time infinitesimally greater than $z_1$ on the Keldysh contour.\\

A similar manipulation can be done in the other four terms arising from the commutator of the interacting Hamiltonian.
By a change of the integration variables and using the relation $\left\langle \hat{\psi}^\dagger (1) \hat{\psi}^\dagger (1')  \hat{\psi} (1') \hat{\psi}_\alpha (1) \right\rangle = - \left\langle \hat{\psi}^\dagger (1') \hat{\psi}^\dagger (1)  \hat{\psi} (1') \hat{\psi}_\alpha (1) \right\rangle$, the last four terms are equal in pair, 
this cancels out the factor $1/2$ and we are left with: 

\begin{align}
\label{eq:mbhc} 
&i \int_\gamma d z_1' \int \int d \bold{x}_1  d \bold{x}_1' \;T_\alpha^*(1') v(1,1') (-i)^2 \left\langle  \hat{\psi}^\dagger_\alpha (1') \hat{\psi}^\dagger (1) \hat{\psi} (1')  \hat{\psi} (1) \right\rangle  +h.c.\\
&= 2 Re \biggl\{i \int_\gamma d z_1'  \int \int d \bold{x}_1  d \bold{x}_1' \;v(1,1')  T_\alpha^*(1') G^{(2)}_{CCC \alpha}  (1',1;1^+,1'^+) \nonumber   \biggr\}
\end{align}
\noindent where we use the definition of the mixed two-particle Green's functions $G^{(2)}_{CCC \alpha}  (1,2;3,4)=(-i)^2 \left\langle \mathcal{T}_\gamma  \hat{\psi} (1)  \hat{\psi} (2) \hat{\psi}^\dagger_\alpha (4) \hat{\psi}^\dagger (3)  \right\rangle $ and $G^{(2)}_{C  \alpha CC} (1,2; 3,4)= (-i)^2 \left\langle \mathcal{T}_\gamma   \hat{\psi} (1) \hat{\psi}_\alpha (2) \hat{\psi}^\dagger (4) \hat{\psi}^\dagger (3) \right\rangle$ together with the relation $G^{(2)}_{C  \alpha CC} (1,1';1'^+,1^+) =  \left[ G^{(2)}_{CCC \alpha}  (1',1;1^+ ,1'^+)\right]^*$.

\noindent By collecting all terms we obtain:

\begin{equation}
\label{eq:currentmix}
\mathcal{\dot E}_C=\sum\limits_\alpha 2 Re \biggl\{ \int d \bold{x} \;h(1) T^*(1) G_{C \alpha}(1^+)  +i \int_\gamma d z'  \int \int d \bold{x}  d \bold{x'} \;v(1;1')  T_\alpha^*(1') G^{(2)}_{CCC \alpha}  (1',1;1^+,1'^+) \biggr\}.
\end{equation}
\end{widetext}

\section{THE MIXED GREEN'S FUNCTIONS}
\label{app:mixed}
We have seen that it is possible to express the energy current flowing through the central interacting region as the sum of two contributions.
These contributions contain the mixed Green's functions accounting for the propagation of both particle in the central region and in the leads.
It is important to point of that such a situation occurs also in the case of the calculation of the particle (charge) current ~\cite{riku2013}.
The usual way to find an expression for the mixed Green's functions $G^{C \alpha}(1;1')$ is based on the equation-of-motion approach ~\cite{Haugbook,stefleebook}. Since the leads are noninteracting, the method allows to write down a closed set of equations and then a general formula for $G^{C \alpha}(1;1')$ in terms of the single particle Green's function. Even though appealing for its simplicity, the equation-of-motion technique  cannot straightforward apply to the mixed two-particle Green's function  $ G_2^{CCC \alpha} (1,2;3,4)$ since the analysis gets quite complicated and a closed set of equations can be found only if one relies on some physical approximations. 
The latter must be chosen consistently with the approximations used for the single-particle Green's function.  \\
Rather than the equations-of-motion technique we find a general expression for both mixed Green's functions by a direct expansion of the $S$-matrix in the interaction picture with respect to the coupling Hamiltonian ~\cite{Haugbook}. Despite the fact that the derivation is somehow more complex for the mixed single-particle Green's function, it is very general and can be extended easily to the mixed n-particle Green's functions.

Let's look at the first term of the energy current, namely the one containing the the single-particle Hamiltonian
$h(1)$.
The key idea of this approach is to express the contour-ordered mixed single-particle Green's function $G_{ C \alpha} (1;1') = -i \left\langle \mathcal{T}_\gamma  \hat{\psi}(1)  \hat{\psi}_\alpha^\dagger (1') \right\rangle$  in terms of the contour-ordered single particle Green's function of the particles in the central region $G_{CC} (1;1')=-i \left\langle \mathcal{T}_\gamma  \hat{\psi}(1)  \hat{\psi}^\dagger(1') \right\rangle$ and the one in the leads $g_{\alpha} (1;1')=-i \left\langle \mathcal{T}_\gamma  \hat{\psi}_\alpha(1)  \hat{\psi}_\alpha^\dagger(1') \right\rangle$.\\
The derivation follows by writing the  Green's function $G_{ C \alpha} (1;1')$ in terms of the interaction-picture operators (denoted by a tilde) with respect to the free Hamiltonians $\hat H_C$ and $\hat H_\alpha$ of both the central region and the terminal $\alpha$. Therefore the evolution operator will be expressed in terms of the coupling Hamiltonian $\tilde {H}_T$ in the interaction picture.
Thus in this picture the mixed single-particle Green's function can be written as:
\begin{widetext}
\begin{equation}
\label{eq:mspgint}
G_{ C \alpha}(1;1') = -i \left\langle \mathcal{T}_\gamma  \hat{\psi} (1)  \hat{\psi}_\alpha^\dagger (1') \right\rangle = -i \left\langle \mathcal{T}_\gamma  \tilde{\psi} (1)  \tilde{\psi}_\alpha^\dagger (1') S  \right\rangle
\end{equation}
where we defined the S-matrix as:
\begin{equation}
S= \sum_{k=0}^\infty \frac{(-i)^k}{k!} \int_\gamma d \bar{z}_1 \dots \int_\gamma d \bar{z}_k  \tilde{H}_T (\bar{1}) \dots \tilde{H}_T (\bar{k})
\end{equation}

By inserting the explicit form of the S-matrix into Eq.~\ref{eq:mspgint} we obtain
\begin{align*}
&-i \left\langle \mathcal{T}_\gamma  \tilde{\psi} (1)  \tilde{\psi}_\alpha^\dagger (1')  \sum_{k=0}^\infty \frac{(-i)^k}{k!}   \int_\gamma \int d \bar{z}_1  d \bar{\bold{x}}_1  \left( \tilde{\psi}^\dagger (\bar{1}) T(\bar{1}) \tilde{\psi}_\alpha (\bar{1})+ h.c. \right) \times \dots \times \int_\gamma d \bar{z}_k \tilde{H}_T (\bar{k})    \right\rangle =\\
&=\int d \bar{1} \sum_{k=0}^\infty \frac{(-i)^k}{k!}  (-i) \left\langle \mathcal{T}_\gamma  \tilde{\psi}_\alpha (\bar{1}) \tilde{\psi}_\alpha^\dagger (1') \right\rangle T(\bar{1})  \left\langle \mathcal{T}_\gamma \tilde{\psi} (1) \tilde{\psi}^\dagger (\bar{1})  \times \dots \times \int_\gamma d \bar{z}_k \tilde{H}_T (\bar{k}) \right\rangle     + \\ &+ \text{$\big ($$k-1$ remaining terms$\big )$}  \\ 
\end{align*}
The key point in the second step is the assumption that the leads are noninteracting allowing to use the Wick's theorem for the $\alpha$-operators. Besides, we use the fact that in the interaction picture the operators $\psi$ and $\psi_\alpha$ are independent and thus the expectations values can be factorized. Finally, by relabeling all integration variables in the remaining $k-1$ it turns out that all these terms are equal and hence we get a factor $k$, therefore:
\begin{align}
\label{eq:mixedspgf}
&G_{ C \alpha}(1;1') =\int d \bar{1} (-i) \left\langle \mathcal{T}_\gamma \tilde{\psi} (1) \tilde{\psi}^\dagger (\bar{1}) \sum_{k=0}^\infty \frac{(-i)^{k-1}}{(k-1)!} \prod\limits_{p=1}^{k} \int_\gamma d \bar z_p \tilde{H}_T (\bar{p})\right\rangle \times T_\alpha(\bar{1})  (-i) \left\langle \mathcal{T}_\gamma  \tilde{\psi}_\alpha (\bar{1}) \tilde{\psi}_\alpha^\dagger (1')  \right\rangle \\
&=\int d \bar{1}\; G (1;\bar{1})  T_\alpha(\bar{1})  g_{\alpha} (\bar{1} ;1')\nonumber  
\end{align}
where in the last line we use $\hat{\psi}_\alpha(1)=\tilde{\psi}_\alpha (1)$, and restore the S-matrix expansion for the single particle Green's function.  \\
The first term in Eq.\ref{eq:currentmix} can be rewritten as:
\begin{align}
\label{eq:nonint}
&\sum\limits_\alpha 2 Re \biggl\{ \int d \bold{x}_1 h(1) T_\alpha^*(1) G_{C \alpha}(1;1^+) \biggr\} = \sum\limits_\alpha 2 Re \biggl\{ \int d \bold{x}_1d\bar{1}\; h(1)   G (1;\bar{1})  \Sigma_\alpha (\bar{1};1^+)  \biggr\}  
\end{align}
where we define the embedding self-energy  $\Sigma_\alpha(1;2) = T_\alpha(1)  g_{\alpha} (1 ;2) T_\alpha^*(2) $. It accounts for the presence of the $\alpha-$ lead.
The second term of Eq.\ref{eq:currentmix} containing the mixed two-particle Green's function $G_{ CCC\alpha }^{(2)}$ can be manipulated in a similar way and rewrite it in terms of the contour-ordered two particle Green's function of the central interacting region $G^{(2)}$:
\begin{align}
&G^{(2)}_{CCC \alpha}  (1',1;1^+,1'^+)=(-i)^2 \left\langle \mathcal{T}_\gamma  \hat{\psi} (1')  \hat{\psi} (1) \hat{\psi}^\dagger_\alpha (1'^+) \hat{\psi}^\dagger (1^+)  \right\rangle = (-i)^2 \left\langle \mathcal{T}_\gamma  \tilde{\psi} (1') \tilde{\psi} (1)  \tilde{\psi}_\alpha^\dagger (1'^+)  \tilde{\psi}^\dagger (1^+) S \right\rangle \\
&=(-i)^2 \biggl\langle \mathcal{T}_\gamma   \tilde{\psi} (1') \tilde{\psi} (1) \tilde{\psi}_\alpha^\dagger (1'^+)  \tilde{\psi}^\dagger (1^+) \times \sum_{k=0}^\infty \frac{(-i)^k}{k!}   \int_\gamma \int d \bar{1}  \left( \tilde{\psi}^\dagger (\bar{1}) T(\bar{1}) \tilde{\psi}_\alpha (\bar{1})+ h.c. \right) \times \dots \times \int_\gamma d \bar{z}_k \tilde{H}_T (\bar{k})  \biggr\rangle \nonumber \\
&= \int d \bar{1}\; (-i)^2 \left\langle \mathcal{T}_\gamma  \tilde{\psi} (1') \tilde{\psi} (1) \tilde{\psi}^\dagger (\bar{1}) \tilde{\psi}^\dagger (1^+) \sum_{k=0}^\infty \frac{(-i)^{k}}{k!}  \prod\limits_{p=1}^k\int_\gamma d z_p \tilde{H}_T (\bar{p}) \right\rangle \times (-i) \left\langle \mathcal{T}_\gamma \tilde{\psi}_\alpha(\bar{1}) \tilde{\psi}_\alpha^\dagger (1'^+) \right\rangle T(\bar{1})  \nonumber \\
&=\int d \bar 1  \; G^{(2)} (1',1;1^+,\bar{1}) T(\bar{1})  g_{\alpha} (\bar{1};1'^+)  \nonumber  
\end{align}
where in the last step we recognize the series expansion in the interaction picture of the two-particle Green's function of the central region  and the single-particle Green's function of the lead $\alpha$.

Inserting this result into the second term of Eq.~\eqref{eq:currentmix} we obtain:

\begin{align}
& i\sum\limits_\alpha  \int d \bold{x}'  d 1 \;v(1',1)  T^*(1') G^{(2)}_{CCC \alpha}  (1',1;1^+ ,1'^+) = i \sum\limits_\alpha \int d \bold{x}'  d 1d\bar{1} \;v(1',1)  T^*(1') G^{(2)} (1',1;1^+,\bar{1}) T(\bar{1})  g_{\alpha \alpha} (\bar{1};1'^+)  \nonumber  \\
&= -i\sum\limits_\alpha  \int d \bold{x}'  d 1 d\bar{1} \; v(1',1)   G^{(2)} (1',1;\bar{1},1^+) T(\bar{1})  g_{\alpha} (\bar{1};1'^+) T^*(1') =\sum\limits_\alpha   \int d \bold{x}'  d 1 d\bar{1} \; \Sigma(1';1) G (1;\bar{1}) \Sigma_{\alpha } (\bar{1}; 1'^+)  \nonumber  \\
\end{align}
where we use $v(1,1')=v(1',1)$, the symmetry relations  of the two-particle Green's function $G^{(2)}(1,2;3,4)=- G^{(2)}(1,2;4,3)$ and  the definition of the embedding self-energy $\Sigma_\alpha  (1;2)$. 

The crucial point in obtaining the above result is to use the relation linking the two-particle Green's function in the interacting region with the many-body self-energy: 
\begin{equation}
\int d \bar{1}\; v(1,\bar{1}) G^{(2)}(1,\bar{1}; 1',\bar{1}^+) = i \int d\bar{1} \; \Sigma(1,\bar{1}) G(\bar{1}, 1').
\end{equation}
which is nothing but the relation which defines the many-body self-energy itself.

Now we can write the final expression in terms of the currents, which in Eq.~\ref{eq:ecurrent2} gives:
\begin{equation}
\frac{d}{dt} \langle \hat{H}_C (t) \rangle =\sum\limits_\alpha\mathcal{J}_C^{(\alpha)}(t)\equiv\sum\limits_\alpha2 Re \biggl \{ \int d \bold{x}_1 d \bar 1 d \bar 2 \biggl[  h(1) \delta(1,\bar  1)  +  \Sigma(1; \bar 1) \biggr] G (\bar  1;\bar  2) \Sigma_{\alpha } (\bar  2, 1^+) \biggr\}_{z_1=t}
\end{equation}

The expression for the variation of the energy has to be compared to the derivative of the total energy:
\begin{equation}
\frac{d}{dt}\mathcal{E}_C(t)= \frac{d}{dz_1} \left[ i \int d \bold{x}_1   d \bar 1 \;\biggl[  h(1) \delta(1,\bar 1)  +\frac{1}{2}\Sigma(1;\bar 1) \biggr] G (\bar 1;1^+)\right]\bigg|_{z_1=t}.
\end{equation}
\end{widetext}

\section{ENERGY VARIATION OF THE CONTACT REGION}
\label{app:coupling}

Here we show how to compute the variation of energy in the contact region. Instead of considering directly the commutator, we exploit some useful identities that make the calculation faster and straightforward. The average variation of energy in the tunneling region then reads: 
\begin{widetext}
\begin{align}
\label{eq:dcplng}
&\frac{d}{dt} \langle \hat{V}^{(\alpha)}_{C} (t) \rangle =2Re \left\{-i \int d \bold{x}_1\; T_\alpha^*(1)\frac{d}{dz_1}G_{C \alpha}(1;1^+)\right\}_{z_1=t}=\nonumber\\
&2Re \left\{\int d \bold{x}_1 d \bar 1\;G(1;\bar 1)\Sigma_{\alpha}(\bar 1;1^+)h_\alpha(1)\right\}_{z_1=t}-2 Re \biggl \{ \int d \bold{x}_1 d \bar 1 d \bar 2 \biggl[h(1) \delta(1,\bar  1)  +  \Sigma(1; \bar 1) \biggr] G (\bar  1;\bar  2) \Sigma_{\alpha } (\bar  2, 1^+) \biggr\}_{z_1=t}\nonumber \\
-&2 Re \biggl \{ \int d \bold{x}_1 d \bar 1 d \bar 2 \;\Sigma_{emb}(1; \bar 1) G (\bar  1;\bar  2) \Sigma_{\alpha } (\bar  2, 1^+) \biggr\}_{z_1=t}\nonumber \\
\end{align}
where we use Eq.~\eqref{eq:mixedspgf} together with the equation of motion for the single particle Green's functions
in the central region and in the leads:
\begin{align}
&i \frac{d}{dz} G(1;1')=\delta(1,1')+h(1)G(1;1')+\int d\bar 1 \;\left(\Sigma(1;\bar 1)+\Sigma_{emb}(1;\bar 1)\right)G(\bar 1;1')\\
&i \frac{d}{dz'} g_{\alpha}(1;1')=-\delta(1,1')-g_{\alpha}(1;1')h(1'),
\end{align}
and where $\Sigma_{emb}(1;1')=\sum\limits_\beta \Sigma_\beta(1;1')$.

It is easy to recognize that the first term is nothing but the opposite of the variation of the energy of the lead $\alpha$:
$\dot{\mathcal{E}}_{\alpha}=\frac{d}{dt} \langle \hat{H}_{\alpha} (t) \rangle $. The second term is $-\mathcal{J}_{C}^{(\alpha)}(t)$ ,namely the opposite of the variation of the energy on the central region due to the coupling with the lead $\alpha$.

The third term is the most interesting one and describes the direct coupling of the lead $\alpha$ with all the others. The coupling is  mediated by the central region and therefore is of the fourth order in the coupling between the central region and the leads.
To see it explicitly, we consider the mixed single particle Green's function $G_{\beta\alpha}(1;1')\equiv-i \left\langle \mathcal{T}_\gamma  \hat{\psi}_\beta(1)  \hat{\psi}_\alpha^\dagger(1') \right\rangle$ and use once again the $S$-matrix expansion:

\begin{align}
\label{eq:mixedtermspgf}
&G_{\beta\alpha}(1;1') \equiv\int d \bar{1}d \bar{2} \;(-i) \left\langle \mathcal{T}_\gamma  \tilde{\psi}_\beta (1) \tilde{\psi}_\beta^\dagger (\bar 1)  \right\rangle T_\beta^*(\bar{1}) \;T_\alpha(\bar{2})  (-i) \left\langle \mathcal{T}_\gamma  \tilde{\psi}_\alpha (\bar{2}) \tilde{\psi}_\alpha^\dagger (1')  \right\rangle \\
&\times(-i) \left\langle \mathcal{T}_\gamma \tilde{\psi} (\bar 1) \tilde{\psi}^\dagger (\bar{2})\sum_{k=0}^\infty \frac{(-i)^{k}}{k!}\prod\limits_{p=1}^k\int_\gamma d z_p \tilde{H}_T (\bar{p})\right\rangle \nonumber \\
&= \int d \bar{1}d \bar{2}\;g_{\beta} (1 ;\bar{1}) T_\beta^*(\bar{1}) G (\bar{1};\bar{2})  T_\alpha(\bar{2})  g_{\alpha} (\bar{2} ;1')\nonumber  
\end{align}

Physically this term represents exactly the scattering of a particle (or a hole) from the lead $\alpha$ to the lead $\beta$
through the central region.
It is now easy to see that the third term in Eq.~\eqref{eq:dcplng} can be written as:

\begin{align}
\label{eq:alphabeta}
\int d \bold{x}_1 d \bar 1 d \bar 2 \;\Sigma_{emb}(1; \bar 1) G (\bar  1;\bar  2) \Sigma_{\alpha } (\bar  2, 1^+)&= \int d \bold{x}_1 d \bar 1 d \bar 2 \;\sum\limits_\beta T_\beta(1)g_{\beta} (1 ;\bar{1}) T_\beta^*(\bar{1}) \;G (\bar{1};\bar{2})  \; T_\alpha(\bar{2})  g_{\alpha} (\bar{2} ;1^+) T_\alpha^*(1) \nonumber\\
&=\int d \bold{x}_1 \sum\limits_\beta T_\beta(1)G_{\beta\alpha}(1;1^+)  T_\alpha^*(1)\nonumber.
\end{align}

It is simple to check that the term $\beta=\alpha$ does not contribute since it is purely imaginary.

\end{widetext}

\section{SELF CONSISTENT GW APPROXIMATION}
\label{app:gw}
\begin{figure}[b]
	\begin{center}
		\includegraphics[width=.5\textwidth]{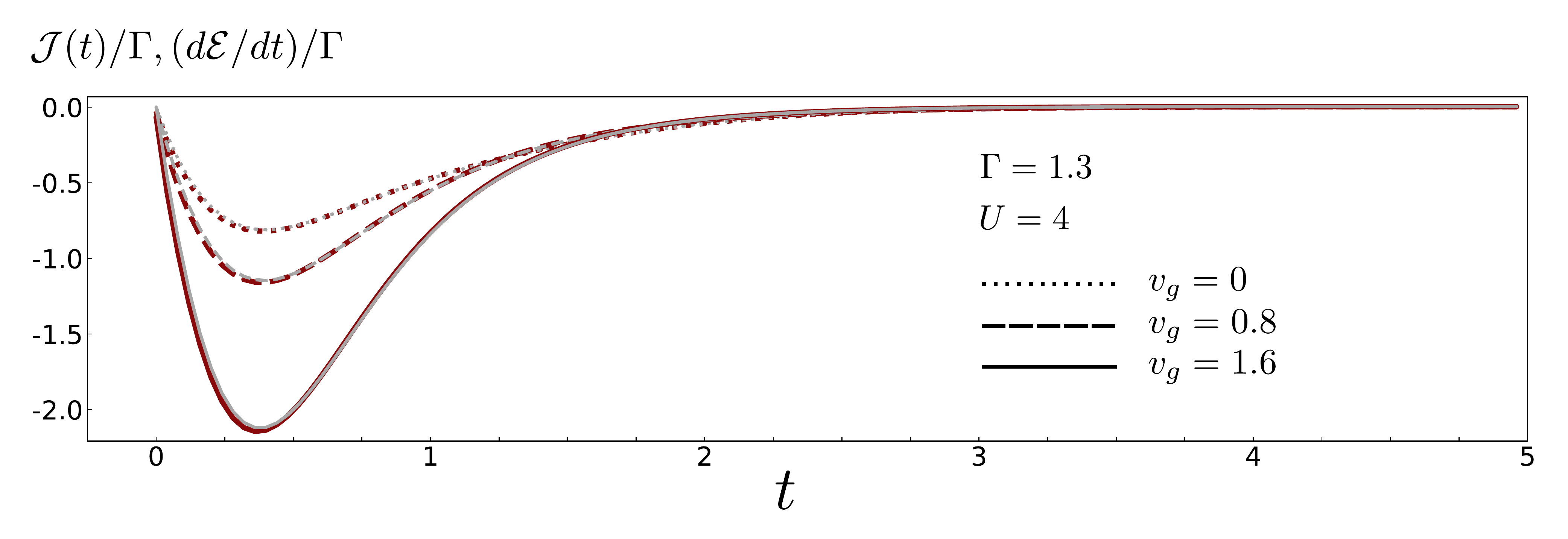}
		\caption{ Energy current $\mathcal{J}_C(t)$ (grey) and rate of change of the total energy $d \mathcal{E}/dt$ (red) as a function of time $t$ in the strong coupling regime $\Gamma=1.3$, charging energy $U=4$ and for different values of the gate-voltage $v_g=0, 0.8, 1.6$ (dotted, dashed, solid).  }
		\label{fig:cons}
	\end{center}
\end{figure}
In this section we recall the main equations to treat the interaction within the GW approximation. It is useful to split the full interaction self-energy into its Hartree and exchange-correlation parts
\begin{equation}
\Sigma(1;2) = \Sigma(1;2)_H +\Sigma(1;2)_{xc}.
\end{equation} 
The Hartree term, or first order polarization diagram, it is local in time and can be written as  $\Sigma(1;2)_H = - i\delta(1;2) \int d 3 v(1;3) G(3;3^+)$. \\
In the GW approximation  the exchange-correlation  part  of  the  self-energy  is  given  as  a product of the Green function $G$ with a dynamically screened interaction $W$:
\begin{equation}
\Sigma(1;2)_{xc} = i W(1;2) G(1;2).
\end{equation}   
The dressed interaction $W$ satisfies the self-consistent equation	
\begin{equation}
W(1;2) =  v(1;2) + \int d3 d4 v(1;3) P(3;4) W(4;2)
\end{equation}  
with $v$ the bare two-body interaction, and $P(1;2) = -i G(1;2) G(2;1)$ the irreducible polarization diagram. 
As it clear from the previous expressions the self-energy happen to be  dependent on the Green’s function and thus it must be determined self-consistently in conjunction with the Dyson equation \cite{talarico2019}. Moreover, the GW self-energy can be express as the functional derivative of a so-called $\Phi$ functional, i.e., $\Sigma_{GW} [G] = \delta \Phi[G]/ \delta G$ . This has been proven to be an effective way to guarantee macroscopic conservation laws \cite{stefleebook}. \\ 
The reliability of the fully self-consistent GW approximation can be easily seen from Fig.~\ref{fig:cons} where we compare the expression obtained for the total current through the interaction region $\mathcal{J}(t)$ in Eq.~\eqref{eq:ecurrent2} alongside the derivative of the total energy in the same region $d \mathcal{E}(t)/dt$, with $\mathcal{E}(t) =\langle \hat{H}_C (t) \rangle = \left[ i \int d \bold{x}_1   d 2 \;\biggl[  h(1) \delta(1,2)  +\frac{1}{2}\Sigma(1,2) \biggr] G (2,1^+)\right]$. It is possible to appreciate the agreement within the numerical accuracy of the two quantities in the strong coupling and for large charging energy at different gate-voltage. The agreement holds also for the other set of parameters considered (not shown) and confirms the validity of the approximation.

\end{document}